%% file: main.tex
\documentclass{article}
\usepackage[T1]{fontenc} % add special characters (e.g., umlaute)
\usepackage[utf8]{inputenc} % set utf-8 as default input encoding
\usepackage{ismir,amsmath,cite,url}
\usepackage{graphicx}
\usepackage{color}
\usepackage{acronym}
\usepackage{booktabs}
\usepackage{amssymb}
\usepackage{pifont}
\usepackage{multirow}

\usepackage{lineno}
\title{Stem-JEPA: A Joint-Embedding Predictive Architecture\\for Musical Stem Compatibility Estimation}
\newcommand{\as}{3mm}
\multauthor
{Alain Riou$^{1,2}$ \hspace{\as} Stefan Lattner$^2$ \hspace{\as} Gaëtan Hadjeres$^3$ \hspace{\as} Michael Anslow$^2$ \hspace{\as} Geoffroy Peeters$^1$}
{ %\bfseries{Michael Anslow$^2$ \hspace{1cm} Geoffroy Peeters$^1$}\\
 $^1$ LTCI, Télécom-Paris, Institut Polytechnique de Paris, France\\
$^2$ Sony Computer Science Laboratories - Paris, France\\
$^3$ Sony AI\\
{\tt\small alain.riou@sony.com}
}

\def\authorname{A. Riou, S. Lattner, G. Hadjeres, M. Anslow, G. Peeters}

\usepackage[bookmarks=false,pdfauthor={\authorname},pdfsubject={\papersubject},hidelinks]{hyperref}
\graphicspath{{figures/}}

\newcommand{\HS}{\mathcal{A}}

\newcommand{\x}{\bold{x}}
\newcommand{\xn}{\bold{x}^{(n)}}
\newcommand{\q}{\bold{q}}
\newcommand{\qn}{\bold{q}^{(n)}}
\newcommand{\z}{\bold{z}}
\newcommand{\zn}{\bold{z}^{(n)}}
\newcommand{\xbar}{\bold{\bar{x}}}
\newcommand{\zbar}{\bold{\bar{z}}}
\newcommand{\ztilde}{\bold{\tilde{z}}}

\acrodef{EMA}{Exponential Moving Average}
\acrodef{JEPA}{Joint-Embedding Predictive Architecture}
\acrodef{ViT}{Vision Transformers}
\acrodef{SSL}{self-supervised learning}
\acrodef{MIR}{Music Information Retrieval}

\DeclareMathOperator*{\argmin}{arg\,min}
\DeclareMathOperator*{\concat}{concat}

\DeclareMathOperator*{\emb}{emb}

\begin{document}

	\maketitle
	\begin{abstract}

        This paper explores the automated process of determining stem compatibility by identifying audio recordings of single instruments that blend well with a given musical context. To tackle this challenge, we present Stem-JEPA, a novel Joint-Embedding Predictive Architecture (JEPA) trained on a multi-track dataset using a self-supervised learning approach.

        %While common self-supervised learning paradigms rely on views (transformed versions) or masked-patch predictions, we propose here a new paradigm based on stem prediction.
        %A given music track is considered as a set of mixed stems (isolated instrument parts) which are therefore considered as sounding well together.
        %Our model comprises two networks (an encoder and a predictor) that are jointly trained to learn semantic representations from an input mix and predict the representations of another stem from the same track that was not included in the initial mix.

        Our model comprises two networks: an encoder and a predictor, which are jointly trained to predict the embeddings of compatible stems from the embeddings of a given context, typically a mix of several instruments.
        
        %This training strategy enables the model to capture global and local semantic information from music clips and to predict a musically coherent representation of a specific instrument at the time of inference.
        %Training a model in such a way allows its use in the estimation of stem compatibility (retrieving, aligning or generating a stem to match a given mix) or downstream tasks (such as genre or key estimation) because the training paradigm forces the model to learn information related to timbre, harmony or rhythm. % proposition de Gaetan et Geoffroy - I like it (Stefan) :)
        Training a model in this manner allows its use in estimating stem compatibility—retrieving, aligning, or generating a stem to match a given mix—or for downstream tasks such as genre or key estimation, as the training paradigm requires the model to learn information related to timbre, harmony, and rhythm.
        
        We evaluate our model's performance on a retrieval task on the MUSDB18 dataset, testing its ability to find the missing stem from a mix and through a subjective user study. We also show that the learned embeddings capture temporal alignment information and,
        %, and for some tasks, we present some failure cases of our model. 
        finally, evaluate the representations learned by our model on several downstream tasks, highlighting that they effectively capture meaningful musical features.
        %\url{https://docs.google.com/document/d/1122GNsj23MprcW4eiQ46fbSmCxJ5u7ebzsi8FRfZOpI/edit?usp=sharing}
	\end{abstract}
	
	\section{Introduction}

Musical stem compatibility indicates the degree to which a stem (i.e., an audio file of a single instrument) fits a given musical context (an audio file of another instrument or a mix of instruments) when played together.
Its automatic estimation can be helpful for stem retrieval, automatic arrangement, or stem generation tasks. The compatibility between stems (or a stem and some musical context) depends on several global factors, such as tonality, tempo, genre, timbre, and singing/playing style. In addition, local features like chords or pitches are crucial to performing temporal alignment between a stem and some musical context.

While initial works have studied musical compatibility between songs based on traditional \ac{MIR} tasks like beat tracking and chord estimation \cite{AutoMashupper, Lee2015}, more modern approaches aim to learn compatibility directly from data using deep neural networks \cite{Huang2021,Chen2020,SampleMatch}.
Using such learning-based approaches extends the notion of compatibility beyond music-theoretical aspects (like tonality and tempo) toward sound-related and expressive characteristics like timbre and playing style.

Moreover, there are potential applications for musical stem generation \cite{DiffARiff, StemGen}, where generators usually require musical context conditioning to produce compatible accompaniments.
With the proposed system, stem representations can be predicted from context information at inference time. This allows training a stem generation model based solely on stem representations, eliminating the need for context/target pairs.

\textbf{Paper proposal and organization.}
% Representation learning has recently gained attention in several domains, including audio and music, due to the possibility of using the learned representations for various downstream tasks, such as classification~\cite{CLMR,MULE}, retrieval~\cite{SampleMatch} and generation~\cite{AudioMAE}.
% In this paper, we propose to use this paradigm to tackle the problem of musical stem compatibility estimation.
% More precisely, we introduce Stem-JEPA, a novel \ac{JEPA} which acts directly on mixtures of musical stems.
In this paper, we introduce Stem-JEPA, a novel \ac{JEPA} which acts directly on mixtures of stems.
It consists of two neural networks, an encoder and a predictor, jointly trained to 
produce representations of a \emph{context} mix and predict representations of a compatible \emph{target} stem.
%produce stem representations that are easy to predict if the stems come from the same music track.
Unlike previous \ac{JEPA}s \cite{IJEPA,M2D}, our approach does not rely on masking in the input space but rather on omitting stems within the process of mixing, and it uses the label of the missing stem for conditioning 
(see section~\ref{sec:method}).

\begin{figure*}
    \centering
    \includegraphics[width=0.9\textwidth]{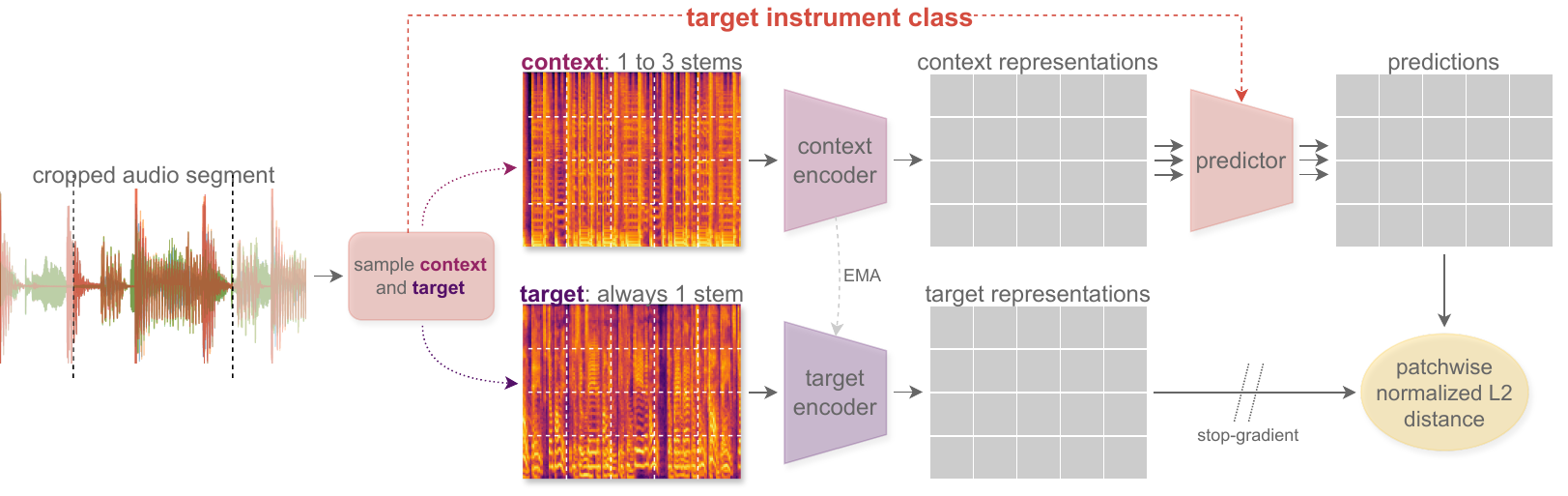}
    \caption{Overview of the Stem-JEPA framework. From an audio clip composed of 4 stems, we first crop a chunk of 8 seconds, then sample the target $\xbar$ (one of the stems) and the context $\x$ (a mix of some of the remaining stems) as described in section \ref{sec:sampling}. They are then converted into Log Mel Spectrograms and passed through the \emph{context} and \emph{target} encoders, respectively. 
    Finally, the \emph{predictor} (conditioned on the target instrument label) is trained so that each of its outputs individually predicts each target representation.}
    \label{fig:main}
\end{figure*}

We assess the performance of Stem-JEPA in a retrieval task and through a subjective evaluation in sections \ref{sec:retrieval} and \ref{sec:user}, respectively. 
Also, we investigate how well the learned representations encode the temporal alignment of stems and mixes (see section~\ref{sec:alignment}). We also perform an analysis showing that key and chord annotations of audio snippets close in the embedding space are musically compatible (see section~\ref{sec:plausibility}). 
Finally, we evaluate the representations produced by our model on various downstream \ac{MIR} tasks (see section~\ref{sec:downstream}).

To facilitate further research in this direction,
we make our code available.\footnote{\url{https://github.com/SonyCSLParis/Stem-JEPA}}

\section{Related work}

\textbf{SSL for representation learning.}
Self-supervised learning (SSL) involves training networks on unlabeled corpora by solving pretext tasks using only the data itself. This paradigm has shown great potential for extracting meaningful representations in various domains~\cite{SimCLR,BYOLA,MULE,BERT}.

A common approach to SSL is contrastive learning~\cite{SimCLR,COLA,CLMR,MULE} or its variants~\cite{VICReg,Wang2020a}. In autoencoders, an encoder and a decoder are jointly trained to learn latent representations from which the original input can be reconstructed~\cite{BERT,MAE}. \ac{JEPA}s are trained to \emph{predict} some target data from context data directly in the representation space~\cite{BYOL,IJEPA}.
%Divising data-specific pretext tasks can strengthen the transferability of the learnt features.
%A frequently used pretext task involves achieving invariance across transformations: we feed a Siamese architecture with two different views of the same data that share underlying semantic information, and the network is then trained to model these similarities. 
%
%Many techniques have been developed to efficiently learn powerful SSL representations \cite{tao2022exploring}: some based on contrastive learning (SimCLR \cite{SimCLR}, CPC \cite{oord2018representation}), masked modeling  (BERT \cite{devlin2018bert}, Masked Autoencoders (MAE) \cite{he2022masked}, data2vec\cite{data2vec}), dual encoders (VICReg\cite{VICReg}, BYOL\cite{BYOL}). Recently I-JEPA\cite{IJEPA} proposed to train the network to predict the masked target \emph{representation} from the unmasked context \emph{representation}, alleviating the need to define the task on data space.

% same as above but no paper names
%Many techniques have been developed to learn powerful representations efficiently \cite{tao2022exploring} some based on contrastive learning \cite{SimCLR,oord2018representation} or masked modeling \cite{devlin2018bert,he2022masked,data2vec}. Recently I-JEPA\cite{IJEPA} proposed to train the network to predict the masked \emph{target representation} from the unmasked \emph{context representation}, alleviating the need to define the task on data space. \TODO{same categories as fig 2 of IJEPA}

\textbf{Joint-Embedding Predictive Architectures.}
A \ac{JEPA} is an architecture composed of two trainable networks: an encoder and a predictor. The model receives a context/target pair as input, passes them through the encoder to create latent representations, and then the predictor is trained to predict the target representation from the context representation.
Pairs can be generated through various data augmentations \cite{BYOL,BYOLA,ATST}, or by masking part of the input, as in data2vec~\cite{data2vec}.
In particular, \ac{JEPA}s do not require negative samples, unlike contrastive approaches~\cite{BYOL}, and enable the model to discard uninformative content given that reconstruction is not required.

%A \ac{JEPA} employs an encoder and a predictor to process a context/target pair from a given input, generating latent representations without requiring negative samples, unlike contrastive methods~\cite{BYOL}. This design allows for the omission of non-essential content since the representations need not be reversible.

To prevent model collapse, it is crucial to block gradients in the non-predictor branch~\cite{SimSiam}, treating its output as the target. Moreover, adopting different but tied encoders for each branch as in Eq.~(\ref{eq:ema})
% (the target encoder's weights being an exponential moving average of the original's),
helps to stabilize training~\cite{SimSiam,BYOL,Tian2021}.
Finally, I-JEPA~\cite{IJEPA} creates pairs through masking and trains the model to predict the representations of small image patches by conditioning the predictor on their positions, allowing the model to grasp local nuances.

%\TODO{add a sentence about local representations with predictor conditioning (i.e. IJEPA)}

%While most methods aim to learn identical representations for different views to capture semantic similarity, I-JEPA~\cite{IJEPA} relies on masking to compute context/target pairs, but here the model is composed of a \ac{ViT}~\cite{ViT} that is trained to predict the representations of small image patches instead of global embeddings, allowing the model to grasp local nuances as well.
%Masked Modelling Duo applies the same approach to spectrograms, achieving notable results in general audio representation learning\cite{M2D}.
%However, applications of \ac{JEPA}s in Music Information Retrieval (MIR) remains unexplored.

\textbf{Learning from separated sources.}
Most SSL approaches, often stemming from the vision domain, have been explored and adapted to the audio domain \cite{COLA,CLMR,MULE,wav2vec,chung2021w2v,AudioMAE,ATST,M2D}.
These works are not specific to musical audio, which is typically composed of several stems providing rich compositional potential for SSL.
In practice, only a few SSL approaches leverage separated stems for tasks such as audio classification~\cite{fonseca2021self}, music tagging~\cite{Garoufis2023a} and beat tracking~\cite{desblancs2023zero}.
Finally, a few works explore modeling the compatibility between stems with applications like automatic mashup creation~\cite{AutoMashupper,Huang2021} and sample or loop retrieval for interactive composition~\cite{Chen2020,SampleMatch}.

\section{Stem-JEPA}\label{sec:method}

 % -------------------------------------
\subsection{Training pipeline}

%Our method is depicted in Figure \ref{fig:main}.
Our method, depicted in Figure \ref{fig:main}, builds upon recent works in \ac{JEPA}s for image and audio representation learning~\cite{IJEPA,M2D}. %\TODO{useless? Geoffroy: non c'est bien}
Given a music track represented as a set of $S$ stems (roughly corresponding to the separated audio sources) $\x_1, \dots, \x_S$, we crop a chunk of 8 seconds.
We then randomly select one of the stems as \textbf{target} $\xbar=x_t$ with $t \in \{1, \dots, S\}$ and use the remaining ones to create a \textbf{context mix}: $\x = \sum_{c \in C} \x_c$ with $C \subset \{1, \dots, S\} \setminus \{ t \}$.

Both $\x$ and $\xbar$ are then converted to Log Mel Spectrograms and divided into a regular grid (over the time and frequency dimensions), leading to $K$ patches. 
The context patches are then fed to a \textbf{context encoder }$f_{\theta}$ to produce patch-wise embeddings $\z = (\z_1, \dots, \z_K)$, where $\theta$ are training parameters.
Similarly, the target patches are fed to a \textbf{target encoder} $f_{\bar{\theta}}$ to produce the patch-wise embeddings
$\zbar = (\zbar_1, \dots, \zbar_K)$.

%In classical \ac{JEPA}s, a \emph{predictor} network would take as input the context representations $\z$ as well as positional encodings indicating the location of the masked patches to predict.
%However, in our scenario masking is done implicitly by picking the stems to use as context and target instead of explicitly dropping patches, so there is no need to model any spatial dependencies.
Finally, the context representations $\z$ are independently fed to a \textbf{predictor} $g_\phi$ (with trainable parameters $\phi$), which is conditioned on the instrument label $l$ of the missing stem by concatenating a learnable embedding $\emb(l)$ to $\z_k$.
% \in \{ \text{bass}, \text{drums}, \text{vocals}, \text{other} \}$ of the missing stem. 
The output of the predictor is therefore the prediction $\ztilde = (\ztilde_1, \dots, \ztilde_K)$, with $\ztilde_k = g_\phi(\concat(\z_k, \emb(l)))$.
% \begin{equation}
%     \ztilde_k = g_\phi(\concat(\z_k, \emb(l)))
%     %\ztilde_k = g_\phi(\z_k, \emb(t)), would be cleaner
% \end{equation}

As in \cite{BYOL,BYOLA,IJEPA}, the parameters $(\theta, \phi)$ of the context encoder and predictor are updated through gradient descent by minimizing the mean squared error $\mathcal{L}(\ztilde, \zbar)$ between the (normalized) predicted and target representations:
\begin{equation}
    \mathcal{L}(\ztilde, \zbar) = \dfrac{1}{K} \sum_{k=1}^{K} \left\| \dfrac{\ztilde_k}{\| \ztilde_k \|} - \dfrac{\zbar_k}{\| \zbar_k \|} \right\|^2,
    %\mathcal{L}(\ztilde, \zbar) = \dfrac{1}{K} \sum_{k=1}^{K} d(\ztilde_k, \zbar_k)
\end{equation}
whereas the parameters of the target encoder $\bar{\theta}$ are updated using an \ac{EMA} of the ones of the context encoder, i.e.,
\begin{equation}
\label{eq:ema}
    \bar{\theta}_i = \tau_i \bar{\theta}_{i-1} + (1 - \tau_i) \theta_i,
\end{equation}
where the \ac{EMA} rate $\tau_i$ is linearly interpolated between $\tau_0$ and $\tau_T$, $T$ being the total number of training steps.

\subsection{Sampling context and target}
\label{sec:sampling}

To avoid training the system on silent target stems or silent context mixes, we first analyze the amplitude content of each of the stems $\x_1, \dots, \x_S$ representing a chunk of a given music track.

Let $\HS \subset \{1, \dots, S\}$ be the indices of active (i.e., non-silent) stems among $\x_1, \dots, \x_S$. We first pick a random index $t \in \HS$ as target\footnote{If $|\HS| < 2$ (a whole chunk is silent or only one active stem), we resample another audio chunk from the same track to prevent having silent context or target.}.
%The target $t$ is selected randomly as one with a high enough amplitude.
%We then randomly select a subset $C \subset \HS \setminus \{ t \}$ from the other non-silent tracks. The size $|C|$ of this subset is sampled uniformly between 1 and the number of remaining non-silent stems $|\HS| - 1$. The prediction task used, in general, more stems in the context than in the target ($|C| > 1$), which makes the task easier for the predictor), but sometimes only one stem ($|C| = 1$). The latter makes the model see individual stems and learn relevant representations for them, which is important since they are then used as targets.
Then, we randomly select a subset $C \subset \HS \setminus \{ t \}$ from the remaining non-silent tracks. The number of stems $|C|$ in this subset is uniformly sampled between 1 and the number of other non-silent stems $|\HS| - 1$. Most of the time, the prediction task incorporates more stems in the context than in the target ($|C| > 1$), simplifying the predictor's task. However, occasionally, the subset consists of only one stem ($|C| = 1$), allowing the model to process individual stems and learn their representations, which is crucial as these are also used as targets.

\subsection{Architecture and training details}
We employ a standard ViT-Base model as the encoder~\cite{ViT}.
Our predictor is a 6-layer MLP with ReLU activations and 1024 dimensions in each hidden layer.
In our ablation studies, the Transformer predictor we use is the same as in \cite{M2D}.

During training, we extract audio chunks of 8 seconds that are converted to log-scaled Mel Spectrograms with 80 mel bins and a window and hop size of 25 and 10 ms, respectively. We use patches of size $16 \times 16$, leading to sequences of $\frac{80}{16} \times \frac{800}{16} = 250$ tokens during training.

We train our model during 300k steps using AdamW~\cite{AdamW}, with a batch size of 256, a base learning rate of $3$e-$4$, and a cosine annealing scheduling after 20k steps of linear warmup. All other hyperparameters are consistent with those used in \cite{M2D}, following their demonstrated effectiveness.
%We refer to our code for the exhaustive list.
Our model is trained for approximately four days on a single A100 GPU with 40 GB of memory.

\subsection{Training data}

We train the model on a proprietary dataset of 20k multi-track recordings of diverse music genres (e.g., pop/rock, R\&B, rap, country) with a total duration of 1350 hours. We use existing instrument annotations to construct four standard categories: Bass, Drums, Vocals, and Other.

% -------------------------------------
\section{Evaluation}

We assess the efficacy of our model to retrieve compatible stems from a given mix through objective and subjective evaluations. We also demonstrate that the learned representations capture local harmonic and rhythmic information. Finally, we show that they also encode high-level features, making them suitable for various \ac{MIR} tasks.%, making them suited for various \ac{MIR} tasks.
% The first task is \emph{stem retrieval} (section~\ref{sec:retrieval}), where the model predicts embeddings for a stem that aligns well with an input mix.
% The second task involves \emph{stem alignment} analysis; here, we examine the cosine distance trends of predicted stem embeddings as the correct stems are temporally shifted relative to the mix (see section \ref{fig:alignment}).
% The third task aims to determine if song fragments that are proximal in the latent space share musically relevant relationships. This is explored by generating a co-occurrence matrix from the key and chord labels of stem embeddings within the same clusters (see section \ref{sec:plausibility}).
% Finally, the fourth task is a benchmark on several \ac{MIR} downstream tasks (section \ref{sec:downstream}). We use the derived embeddings with a linear probe classifier to estimate music attributes such as genre, tempo, or key. This evaluation helps identify the specific musical content captured in our embeddings.\TODO{remove stuff here}

% -------------------------------------
\subsection{Stem retrieval task}\label{sec:retrieval}

Given an input audio, our model predictor has been trained to output a latent representation of a stem such that this stem would fit well with the input audio.
To evaluate the performance of our model, we construct a retrieval task in which, given an existing music track, the model should be able to predict the representation of one stem given the mix of the others. 

% -------------------------------------
\subsubsection{Experimental setup}\label{sec:setup}

For evaluation, we used the MUSDB18 dataset~\cite{MUSDB18}, which contains $N = 150$ tracks $\x^{(1)}, \dots, \x^{(N)}$, each track $\xn$ being composed of $S = 4$ stems $\xn_1, \dots, \xn_S$ (vocals, bass, drums, other).
This allows a total of $N \times S = 600$ runs.
For any individual stem $\xn_s$, define $\xn_{\neg s}$ the mix containing all stems from $\xn$ except $\xn_s$. We aim to predict the embedding of the individual stem $\xn_s$ from the one of the mix $\xn_{\neg s}$.
We compute and average (over time) the patch-wise representations of all stems $\xn_s$. 
It gives us a \emph{reference set} $\bold{Z} = \{ \zn_s \}$, with $\zn_s$ being the embedding of $\xn_s$.
Then, we encode all mixes $\xn_{\neg s}$, pass the resulting representations through the predictor conditioned on $s$, and average (over time and frequency) the result to get a \emph{query} embedding $\qn_s$.
In other words, $\qn_s$ is the prediction of (the embedding of) the missing instrument $\xn_s$ from the remaining ones $\xn_{\neg s}$.
We therefore test if the actual embedding $\zn_s$ is among the nearest neighbors of $\qn_s$ in the reference set $\bold{Z}$.

\textbf{Metrics.} We measure the model performance using two metrics.
The \textit{Recall at $K$} (R@$K$)
%\footnote{This metric is widely used in the field of multimodal self-supervised learning~\cite{CLIP,CLAP}.}
measures the proportion of relevant items successfully retrieved among the top $K$ nearest neighbors. We consider here $K \in \{1, 5, 10\}$.

The \textit{Normalized Rank}~\cite{SampleMatch} of a query $\qn_s$ is defined as the rank of the ground-truth $\zn_s$ in the sorted list of distances $\{ d(\qn_s, \z) \}_{\z \in \bold{Z}}$, normalized by the length of the list (here 600) to get a value in $[0, 1)$.
For example, a mean Normalized Rank of 5\% means that the actual embedding $\zn_s$ is, on average, within the 5\% nearest neighbors from the prediction $\qn_s$. For each model, we report mean and median Normalized Ranks.

% -------------------------------------
\subsubsection{Results}

\input{tables/retrieval}

%\input{tables/instruments}

%In this part, we want to observe to what extent our model is able to capture global musical information. We therefore average representations to get a single global embedding per stem and per track, for a total of $N \times S = 600$ pairs of representations and predictions.

The results are shown in Table~\ref{tab:retrieval} under the row "MLP w/ cond." Our model achieves a R@1 of 33\%, and in half of the cases, the correct stem is within the top 0.5\% of nearest neighbors (median Normalized Rank is 0.5\%). Moreover, the median rank consistently outperforms the mean rank, indicating the presence of outliers with very high ranks.

We also include in Table~\ref{tab:retrieval} results for scenarios without predictor conditioning during training (row "MLP w/o cond.") and when using a Transformer instead of an MLP for the predictor. In both cases, the performance drops substantially, emphasizing the importance of conditioning for the retrieval task. When using an MLP instead of a Transformer, the encoder must capture global information because the MLP cannot infer it, which leads to more informative embeddings.

Finally, we compare our model with AutoMashupper~\cite{AutoMashupper}, which is, to the best of our knowledge, the only openly available work on compatibility estimation. We use their ``mashability'' measure as a similarity metric to compute the retrieval performances. Note that this metric involves beat tracking and chord detection, making it unsuited for vocals and drum stems, respectively. Therefore, the performance of this method on the retrieval task is relatively weak.

% -------------------------------------
\subsubsection{Influence of the instrument class}\label{sec:failures}
\begin{figure}
    \centering
    \includegraphics[width=0.45\textwidth]{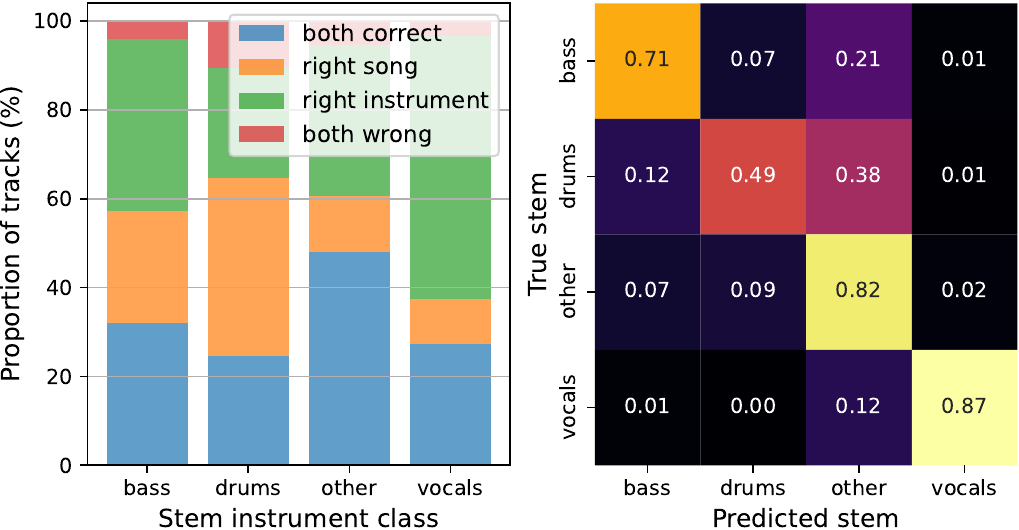}
    \caption{Analysis of the closest embedding $\z^*$ for all queries $\q$ from the MUSDB18 dataset~\cite{MUSDB18}. 
    Left: Categories of failures for each instrument (same song but wrong instrument, the opposite, or both wrong). 
    Right: confusion matrix between conditioning instruments and retrieved instruments.}
    \label{fig:failures}
\end{figure}
% In Table~\ref{tab:instruments}, we detail the performance for each instrument.
% The most noticeable result is the higher performance in predicting ``other'' (R@$1=41.3 \%$) compared to ``drums'' (R@$1=15.3 \%$). A possible reason for this is that the representations we used for this task are averaged over time which could eliminate some temporal information. However, it is more likely that there are simply more possible candidate drum patterns that actually fit a given mix resulting in more close neighbors within which it is harder to detect the ground truth. 

%- normalized rank is lower overall than for any individual stem, which means that the model usually predicts the right kind of instrument?
%- opposite for recall, recall on average is much better than overall, what does it mean?

To get a better understanding of the failure modes of our model and the disparities between the different instruments, we study the nearest neighbor $z^* = \argmin_{\z \in \bold{Z}} d(\q, \z)$ for all queries $\q$ from MUSDB18.
This analysis, detailed in Figure \ref{fig:failures} (left), categorizes $z^*$ into four groups: ``both correct'' where the model predicts the correct instrument from the correct song, ``right instrument'' where the correct instrument is predicted but from a different song, ``right song'' where the model predicts the wrong instrument class but from the correct song, and ``both wrong''. Additionally, Figure~\ref{fig:failures} (right) displays a confusion matrix for the instruments.
%\TODO{Alain, maybe we shouldn't use "right and wrong" for songs in the figure, but rather same/different (maybe also for instruments, even though there right/wrong fits better.) Geoffroy: moi j'aime bien correct/wrong, au moins c'est clair; sinon il faut rappeler qu'on cherche les "same". }

A noticeable result is that the retrieval performances vary a lot between the different instruments, especially between ``drums'' (R@$1 \simeq 25\%$) and ``other'' (R@$1 \simeq 45\%).$\footnote{The proportion of ``both correct'' samples is exactly the Recall at 1.}
A plausible explanation is that there are simply more possible candidate drum patterns that actually fit a given mix, resulting in closer neighbors within which it is harder to detect the ground truth. 

Additionally, we can see that the ``both wrong'' scenario is quite uncommon.
However, for bass and drums in particular, we predict another instrument (but for the correct song) in more than 25\% of the cases.
The confusion matrix shows that the category that mostly causes this failure is ``other''.
A reason is probably that ``other'' is a broad and ill-defined set of instruments that could arguably overlap with bass or drums (e.g., choirs, synth bass, xylophone...).

% ~~~~~~~~~~~~~~~~~~~~~~~~~~~~~~~~~~~~~
\subsection{User study}\label{sec:user}

\begin{figure}
    %\centering
    \includegraphics[width=0.48\textwidth]{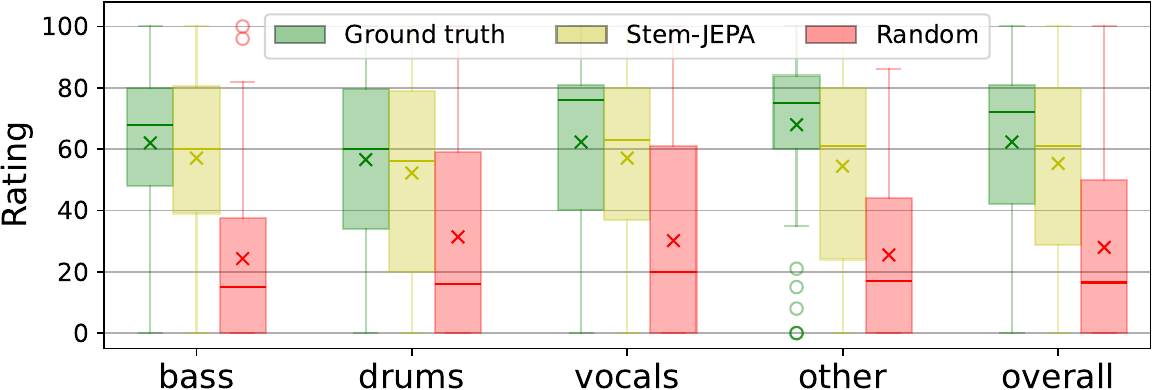}
    \caption{Box plot of the listening test for the different instrument classes. The $\times$ represents the mean of the data.}
    \label{fig:user}
\end{figure}

In section \ref{sec:retrieval}, we utilize the compatibility of a mix and a stem from the same song to assess the retrieval performance of our model. However, it is plausible that the dataset also includes compatible stems originally part of different songs. To evaluate our model's ability to retrieve these compatible yet non-original stems, we conduct a listening test, focusing on retrieving instruments that are not present in the query mix (green segment in Figure \ref{fig:failures}).

%We use the notations introduced in section \ref{sec:retrieval}.
% For each trial, the user first listens to a reference context mix with one missing stem $\xn_{\neg s}$, followed (in random order) by the actual missing stem $\xn_s$, the one retrieved by our system ($\x^{(n*)}_s$ s.t. $\z^{(n*)}_s = \argmin_{\z \in \bold{Z}} d(\qn_s, \z)$), and a random one, but with the same instrument class as $\xn_s$.
For each trial, the user first listens to a query mix with one missing stem, followed (in random order) by the actual missing stem, the one retrieved by our system, and a random one, but with the same instrument class as $\xn_s$.
They are then asked to rate (from 0 to 100) the three proposed stems' compatibility with the reference mix.\footnote{Since we already test temporal alignment and tonality in sections \ref{sec:alignment} and \ref{sec:plausibility} respectively, participants are explicitly instructed to rather concentrate on genre, timbre, and playing/singing style in this study.}

The mixes and stems are 16-second chunks from the MUSDB18 dataset~\cite{MUSDB18}, randomly cropped to 10 seconds during the test to prevent listeners from relying on temporal alignment for rating.
We conduct our study on the Go Listen platform~\cite{GoListen}.
Our test comprises 60 trials, and each user has to answer 12 of them (3 for each instrument class).
We had 23 participants, 20 of whom had musical experience (11 for at least 10 years).

\textbf{Results.}
The listening test results are depicted in Figure \ref{fig:user}. While the ratings for the stems retrieved by our model are slightly lower than those for the ground truth on average, they are substantially higher (approximately double) than the ratings of random samples. This highlights the ability of Stem-JEPA to retrieve stems compatible with the context mix.

However, we find some disparities between instrument classes. For example, the ratings for drums between the ground truth and our model's suggestions are very close, whereas they are more different for the ``other'' category. Also, the variance of ratings is higher in ``drums,'' hinting at a generally higher compatibility of drums with any context.
%Additionally, the task's difficulty is evidenced by the presence of numerous 
%outliers, as indicated by the length of the whiskers.
Finally, the length of the whiskers and the difference between the mean and median reveal significant disparities between users and samples, indicating the high subjectivity and difficulty of musical compatibility estimation.
% \begin{itemize}
%     \item results are shown on Figure \ref{fig:user}
%     \item ratings of the stems retrieved by our model, while not as good as the ratings of the ground truth, are significantly better than the ones of the random samples, highlighting the capabilities of Stem-JEPA to retrieve stems that are compatible with the context mix
%     \item some disparities between instrument classes: ratings between ground truth and our model are very close for drums but more spread for the ``other'' category, which makes sense since this category is way more discriminative
%     \item finally, the task is quite hard and there are a lot of outliers, as shown by the length of the whiskers.
% \end{itemize}

% -------------------------------------
\subsection{Stem alignment analysis}\label{sec:alignment}

\begin{figure}
    %\centering
    \hspace{0.025\textwidth}
    \includegraphics[width=0.4\textwidth]{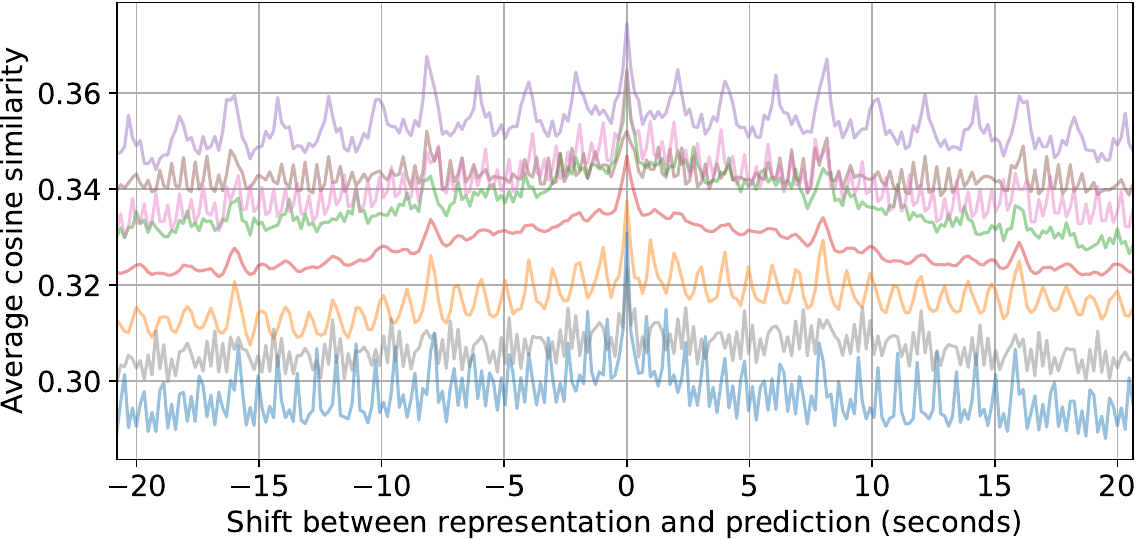}
    \caption{Average pairwise cosine similarity between embeddings and predictions across various temporal shifts. Each curve corresponds to a different track.}
    \label{fig:alignment}
\end{figure}

In this section, we assess the model's ability to evaluate the alignment between stems and mixes by temporally shifting them relative to each other. Our primary metric for this evaluation is the cosine similarity between learned embeddings and their predictions at various offsets, reflecting the local temporal features captured by the model.

Contrary to our previous approach that utilized embeddings averaged over time, here we retain the temporal sequence of the embeddings. We concatenate embeddings in the frequency dimension and stack them in the time dimension, maintaining a resolution of one embedding per 160 milliseconds of audio. We denote the representation of the \(i\)-th patch in stem \(\xn_s\) as \(\zn_s[i]\) and its corresponding predicted output conditioned on the mix $\xn_{\neg s}$ as \(\qn_s[i]\).

We evaluate the fidelity of these embeddings by examining how the cosine similarity between \(\zn_s[i]\) and \(\qn_s[(i+j) \% M]\) evolves with varying \(j\), the temporal offset. The formulation is given by:
\begin{equation}
    s(\z, \q, j) = \dfrac{1}{MS} \sum_{s=1}^S \sum_{i=1}^M \langle \z_s[i], \q_s[(i+j) \% M] \rangle
\end{equation}
where \(M\) is the total number of embeddings in sequence \(\z\), \(S\) is the number of stems, and \(j\) represents the shift index.

The local nature of the information captured by the embeddings is reflected in how the cosine similarity \( s(\z, \q, j) \) changes with different temporal offsets \( j \). Specifically, if the embeddings predominantly contained global information, \( s \) would remain relatively constant across shifts. Conversely, a sharp peak in similarity at \( j = 0 \), followed by a rapid decrease, suggests that the embeddings are rich in local information and less information is shared between adjacent frames.

From our analysis of tracks from the MUSDB18 dataset (8 of them being displayed in Figure \ref{fig:alignment}), we first observe that \( s(\z, \q, j) \) always remains relatively high\footnote{As a reference, the average cosine similarity between random representations and predictions is approximately 0.17.}, which indicates that the embeddings contain global information.
We, however, observe a peak at $j = 0$, underlining the presence of local details that are temporally aligned.

We also observe periodic patterns in the curves, highlighting the model's capacity to capture temporal structures (e.g., beats and bars).
Finally, we observe smaller peaks every 8 seconds, the duration of the chunks used for computing the embeddings, which implies that embeddings also capture global position information. This behavior could potentially be avoided by replacing the absolute positional encodings in our encoder with other variants.

% - whatever the shift, the cosine similarity remains high (average of 0.17 for random embeddings), so it contains global information
% - a sharp peak at $j = 0$, so embeddings also encode local information
% - periodic patterns: embeddings capture temporal structures (beats and bars)
% - artifacts every 8 seconds (i.e. size of the chunks during training), means that embeddings encode global position information. this is undesirable and suggests using different positional embeddings could be beneficial

% From our analysis of 15 tracks from the MUSDB18 dataset, displayed in Figure \ref{fig:alignment}, we observe that the cosine similarity generally surpasses that of random embeddings, which implies some level of temporal (alignment) clustering. The similarity is maximal when there is no shift (\( j = 0 \)) and diminishes progressively, underscoring the presence of local details that are temporally aligned. Additionally, the fluctuating pattern of similarity decay, showing pseudo-periodicity, highlights the model's capacity to capture temporal structures (e.g., beats and bars), albeit with variations across different tracks.

An interactive version of Figure \ref{fig:alignment} with audio examples is provided on the accompaniment website.\footnote{\url{https://sonycslparis.github.io/Stem-JEPA}}

\begin{figure}
    %\centering
    \includegraphics[width=0.45\textwidth]{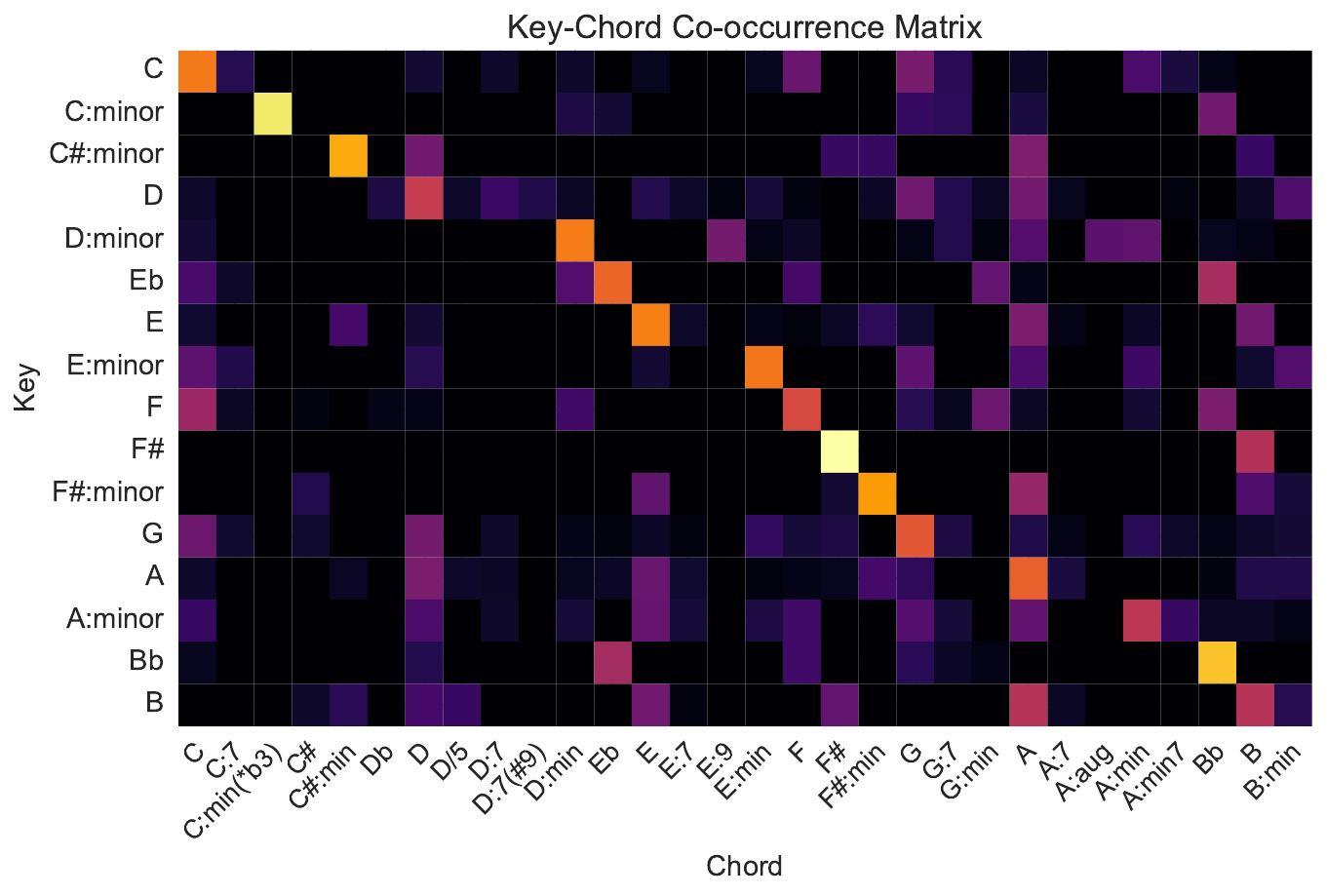}
    \caption{Key/Chord co-occurrence matrix between segments within the same clusters.}
    \label{fig:cooccurrence}
\end{figure}

\subsection{Musical plausibility}\label{sec:plausibility}

We utilize key and chord annotations from Isophonics\footnote{\url{http://isophonics.net/}} for 174 Beatles songs to label all patch embeddings. We then conduct $k$-means clustering on the latent space with \(k=32\) clusters. Within each cluster, we calculate the co-occurrence of all key and chord pairs and aggregate these counts across all clusters. To visualize these relationships, we display in Figure \ref{fig:cooccurrence} a co-occurrence matrix for the keys and chords that appear in the top 80 most frequent combinations, considering all possible pairs for counting, not just the most common ones.
%Figure \ref{fig:cooccurrence} shows the resulting co-occurrence matrix.

The matrix reveals that pairs close in the latent space often share significant musical relevance. The highest occurrences typically connect a key with its tonic (e.g., E/E), and prominently with its subdominant and dominant (e.g., C/F, G or D/G, A). Such patterns indicate that the embeddings capture meaningful tonal relationships.

% -------------------------------------
%\newpage
\subsection{Benchmark on downstream tasks}\label{sec:downstream}

Lastly, we investigate the musical features encoded in the representations learned by our model. We hypothesize that the encoder captures shared musical information among different stems of the same track, such as rhythm or harmony, to aid the predictor. To verify this, we evaluate it on several downstream classification tasks, a standard protocol for representation learning methods~\cite{DBLP:conf/nips/TurianSKRSSMTVM21,MULE,M2D,MARBLE}.

% -------------------------------------
\subsubsection{Experimental setup}

Our experimental setup follows the constrained track of the MARBLE benchmark~\cite{MARBLE}.
%We train an MLP classifier on top of the representations returned by our frozen encoder.
% Specifically, each audio sample from the downstream dataset is processed by the encoder, with its patch-wise outputs concatenated and averaged along the frequency and time dimensions to obtain a single 3840-dimensional global embedding per audio, as done in \cite{M2D}. This embedding is then passed through an MLP with 512 hidden units and a softmax layer, which maps the representations to the appropriate classes. The network is trained in a supervised manner by minimizing the cross-entropy between the predicted distribution and the ground truth labels.
% More precisely, each audio of the downstream dataset is fed to the encoder, whose patch-wise outputs are then concatenated and averaged along the frequency and time dimension respectively, to get a single 3840-d global embedding per audio, as e.g. in \cite{M2D}. Then a neural network composed of an MLP with 512 hidden units and a softmax maps those representations to the appropriate number of classes and is trained in a supervised way by minimizing the cross-entropy between the predicted distribution and the ground truth label.
Each audio sample is processed by the \emph{frozen} encoder, and its patch-wise outputs are concatenated and averaged along frequency and time dimensions to produce a 3840-dimensional global embedding, following \cite{M2D}. These embeddings are passed through an MLP with 512 hidden units and a softmax layer, which is trained by minimizing the cross-entropy between the predicted distribution and the ground truth labels.
%We refer to \cite{MARBLE} for the detailed set of hyperparameters.

\textbf{Downstream tasks.} To validate our hypothesis, we focus on global musical features that are shared among the different stems of a track, namely tagging, key, and genre estimation.
Additionally, we include an instrument classification task to observe whether the encoder preserves stem-specific information. For facilitating comparisons to existing work, we also pick our downstream tasks from the MARBLE benchmark~\cite{MARBLE}. The full list of datasets and associated tasks is depicted in Table \ref{tab:tasks}.
%For a more in-depth description of the datasets, tasks, and corresponding metrics, we refer the reader to \cite{DBLP:conf/nips/YuanMLZCYZLHTDW23}. 

\begin{table}
\centering
\caption{Datasets used for downstream tasks.}
\label{tab:tasks}
\footnotesize{
\begin{tabular}{lcl}
    \toprule
    Dataset & classes & Task \\
    \midrule
    Giantsteps (GS) \cite{Giantsteps} & 24 & Key detection \\
    GTZAN \cite{GTZAN} & 10 & Genre classification \\
    %GTZAN\small{tempo} \cite{GTZAN} &  & Tempo estimation \\
    MagnaTagATune (MTT) \cite{MTT} & 50 & Tagging \\
    NSynth \cite{NSynth} & 11 & Instr. classification \\
    \bottomrule
\end{tabular}
}
\end{table}

\textbf{Baselines.} We compare our model to two variants: one trained with a Transformer as predictor, and one without conditioning, as in section \ref{sec:retrieval}.
%\footnote{Note that the predictor was used for training the model but is discarded for this experiment. We only use the outputs of the frozen encoder.}
In addition, we include the two top-performing models from \cite{MARBLE} in the considered tasks, namely MULE~\cite{MULE} and Jukebox-5B~\cite{DBLP:journals/corr/abs-2005-00341} as references.
MULE~\cite{MULE} is an SSL model based on SF-NFNet-F0~\cite{DBLP:conf/icassp/WangLWRSBJADCO22} trained by contrastive learning on the MusicSet dataset (117k hours), while Jukebox~\cite{DBLP:journals/corr/abs-2005-00341} is a huge music generation model trained using codified audio language modeling on 1.2 million songs.

For a more in-depth description of the hyperparameters, datasets, tasks, and corresponding metrics, we refer the reader to \cite{MARBLE}. 

% \textbf{Tempo estimation.}
% We add tempo estimation to the tasks of MARBLE.
% It is evaluated using accuracy1 (correct tempo considering a 4\% tolerance) and accuracy2 (the same but allowing "octave" errors). 
% It is analogous to Raw Chroma Accuracy in the field of pitch estimation.
% - however, we do not train the linear classifier to predict tempo classes directly
% - In fact, for a given track $\x$ and tempo $t$, we compute the label $y$ as:
% \begin{equation}
% y = \texttt{round} \left( \dfrac{\log t}{2 \log (1.04)} \right)
% \end{equation}
% and then train a linear classifier on pairs $(\z, y)$ as for other downstream tasks, where $\z$ denotes the average-pooled representation of $\x$.

% The choice of the mapping comes from the definition of Accuracy1 in the field of tempo estimation, which usually has a tolerance of 4\%. With this mapping, different tempi with a relative difference of 4\% are likely to be mapped to the same $y$. This is analogous to semitones in the field of fundamental frequency estimation, and we found in our previous experiments that applying this mapping to tempo estimation significantly improved the downstream performances.\TODO{looks like a lot for no great results, maybe drop everything about tempo, or at least these explanations}

% -------------------------------------
\subsubsection{Results}

\input{tables/downstream}

% \begin{figure}
%     \includegraphics[width=0.42\textwidth]{figs/cm}
%     \caption{Confusion matrix of the downstream predictions on the test set of GTZAN. \TODO{should be removed right?}}
%     \label{fig:enter-label}
% \end{figure}

%\TODO{Predictor less important than for retrieval
%- say we are slightly below baselines but significant less data (but separated stems so it's a bit cheating)}

The performances of our model on downstream tasks are provided in Table~\ref{tab:downstream}.
First, we observe that the choice of the predictor used for training, while extremely influencing for retrieval tasks, has little effect on the downstream performances of our encoder, apart from key detection on Giantsteps, for which the model trained with a Transformer predictor clearly outperforms the others.
The performances on NSynth also reveal that our model does not only capture information shared between stems but also stem-specific features.
Surprisingly, this holds even without conditioning the predictor during training, and more generally, not conditioning the predictor improves performance on most downstream tasks.

% - predictor used during training has less
% - transformer = MLP
% - conditioning slightly hurts

% First, we observe that, among our models, the one whose predictor is an MLP conditioned on the instrument target label usually outperforms the two others on all tasks, except for genre classification, for which the difference is marginal ($-0.4\%$). This is consistent with our previous results on retrieval tasks, though less pronounced (see Table \ref{tab:retrieval}). It confirms that letting the encoder handle all time-frequency dependencies and conditioning the predictor both lead to better representations.

% We also observe that our model does not only capture shared features between stems but also stem-specific timbre information, as the results on NSynth reveal ($73.5\%$). Interestingly, this holds even without conditioning.

We also compare our model to state-of-the-art works in music representation learning. Our performances are on par with baselines for two tasks (MTT and NSynth) but significantly lower on Giantsteps and GTZAN, despite being much better than random guessing. Considering the limited quantity of training data compared to the baselines (about 100 times less), these results suggest that our method is promising for music representation learning but that further efforts have to be made to make it competitive with current state-of-the-art approaches in this field.

% -------------------------------------
\section{Conclusion}

In this study, we introduce a novel SSL paradigm based on stem prediction for musical stem compatibility estimation through the prism of representation learning.
Our results show promising performances for retrieval applications and also indicate that the learned representations are localized, suggesting that they could also be valuable for music generation and possibly automatic arrangement. Additionally, these representations are musically meaningful and demonstrate linear separability for various Music Information Retrieval tasks.
%While common self-supervised learning paradigms rely on views (transformed versions) or masked-patch predictions, we introduce a new SSL paradigm based on stem prediction.

Moreover, our model is, to the best of our knowledge, the first use of the predictor component of Joint-Embedding Predictive Architectures (JEPAs) during inference. Employing JEPAs to model compatibility instead of similarity, with appropriate conditioning, may open up possibilities in various fields beyond music. 

Nevertheless, our study is not without its limitations. In particular, self-supervised learning usually benefits from very large corpora of training data; however, accessing large datasets of separated stems is challenging, though advancements in source separation technology may alleviate some of these issues.
Finally, restricting the analysis to four instruments, while standard in source separation, currently limits the generalizability of our findings. Ideally, extending the predictor to accommodate any instrument would prevent the failure cases illustrated in section \ref{sec:failures} and enhance the model's utility, representing an exciting direction for future research.

%\TODO{check references are clean and shorten them (authors et al., remove webpages, remove editors)}

\section{Acknowledgments}

This work has been funded by the ANRT CIFRE convention n°2021/1537 and Sony France. This work was granted access to the HPC/AI resources of IDRIS under the allocation 2022-AD011013842 made by GENCI. We would like to thank Cyran Aouameur and Marco Comunità for their relevant suggestions, as well as Amaury Delort for his help with data. Finally, we would like to thank the reviewers and meta-reviewer for their valuable comments.

% For bibtex users:
\bibliography{AJEPA}
%\printbibliography

% For non bibtex users:
%\begin{thebibliography}{citations}
% \bibitem{Author:17}
% E.~Author and B.~Authour, ``The title of the conference paper,'' in {\em Proc.
    % of the Int. Society for Music Information Retrieval Conf.}, (Suzhou, China),
% pp.~111--117, 2017.
%
% \bibitem{Someone:10}
% A.~Someone, B.~Someone, and C.~Someone, ``The title of the journal paper,''
%  {\em Journal of New Music Research}, vol.~A, pp.~111--222, September 2010.
%
% \bibitem{Person:20}
% O.~Person, {\em Title of the Book}.
% \newblock Montr\'{e}al, Canada: McGill-Queen's University Press, 2021.
%
% \bibitem{Person:09}
% F.~Person and S.~Person, ``Title of a chapter this book,'' in {\em A Book
    % Containing Delightful Chapters} (A.~G. Editor, ed.), pp.~58--102, Tokyo,
% Japan: The Publisher, 2009.
%
%
%\end{thebibliography}

%

\end{document}

%% file: tables/retrieval.tex
\begin{table}
    \centering
    \caption{Influence of the design of the predictor on the retrieval performances. All metrics are in percentages.}
    \label{tab:retrieval}
    {\footnotesize
        \begin{tabular}{lccccccc}
            \toprule
            & \multicolumn{3}{c}{Recall $\uparrow$} &  & \multicolumn{2}{c}{Normalized Rank $\downarrow$} \\
            %\cline{3-3} \cline{5-5} \cline{7-8} \cline{10-11} \cline{13-13} \cline{15-15}
            Model &
            R@1 & R@5 & R@10 &  & mean & median \\
            \midrule
            MLP	w/ cond.     & \textbf{33.0} & \textbf{63.2} & \textbf{76.2} &  &  \textbf{2.0} &  \textbf{0.5} \\
            %\midrule
            MLP w/o cond.    & 28.2 & 58.0 & 69.2 &  &  3.3 &  0.7 \\
            Transformer	     &  5.2 & 17.5 & 25.7 &  & 12.1 &  6.0 \\
            \midrule
            AutoMashupper &
            1.0 &  8.8 & 15.5 &  & 29.1 & 19.5 \\
            % Contrastive &
            % 6.2 & 74.2 & 85.5 &  &  1.7 &  0.5 \\
            \bottomrule
        \end{tabular}
    }
\end{table}

%% file: tables/downstream.tex
\begin{table}
    \centering
    \caption{
        Influence of the predictor architecture on the performances of Stem-JEPA on various downstream tasks, and comparison with existing baselines.
    }
    \label{tab:downstream}
    {\footnotesize
        \begin{tabular}{lccccc}
            \toprule
             & GS & GTZAN & \multicolumn{2}{c}{MTT} & NSynth \\
            Model & Acc\textsuperscript{refined} & Acc & ROC & AP & Acc \\
            \midrule
            MLP w/ cond.        &
            40.2 & 68.6 & 89.9 & 42.8 & 73.5 \\
            %MLP 2 layers         & 49.4 & 21.7 / 24.1 & 65.5 &  \\
            MLP w/o cond. &
            36.8 & 72.5 & 90.1 & \textbf{42.9} & \textbf{75.0} \\
            Transformer &
            46.0 & 68.1 & 90.0 & 42.7 & 73.3 \\
            % Contrastive &
            % 61.3 & 68.5 & 89.1 & 40.7 & 75.3 \\
            \midrule
            MULE \cite{MULE} &
            \textbf{64.9} & 75.5 & 91.2 & 40.1 & 74.6 \\
            %MULE &
            %50.8 & 68.6 & 90.5 & 38.9 & 71.4 \\
            Jukebox \cite{Jukebox} &
            63.8 & \textbf{77.9} & \textbf{91.4} & 40.6 & 70.4 \\
            \bottomrule
        \end{tabular}
    }
\end{table}